\documentclass[twocolumn,pre,amsmath,amssymb,showpacs,
superscriptaddress,floatfix]{revtex4}

\usepackage{color}
\usepackage{hyperref}
\usepackage{graphicx}
\DeclareMathOperator{\Tr}{Tr}  
\begin{document}

\title{Pressure induced orientational glass phase in molecular
para-hydrogen. }

\author{T.I. Schelkacheva}
\affiliation{Institute for High Pressure Physics, Russian Academy of Sciences, Troitsk 142190,
Moscow Region, Russia}

\author{E.E. Tareyeva}
\affiliation{Institute for High Pressure Physics, Russian Academy of Sciences, Troitsk 142190,
Moscow Region, Russia}

\author{N.M. Chtchelkatchev}
\affiliation{Institute for High Pressure Physics, Russian Academy of Sciences, Troitsk 142190,
Moscow Region, Russia}

\date{\today}

\begin{abstract}
We propose a theoretical description of possible orientational glass
transition in solid molecular para-hydrogen and ortho-deuterium  under
pressure supposing that they are mixtures of $J=0$ and $J=2$ states of
molecules. The theory uses the basic concepts and methods of standard
spin-glass theory. We expect our orientational glass to correspond the II'
phase of the high pressure hydrogen phase diagram.
\end{abstract}

\pacs{}

\maketitle

\section{Introduction\label{Sec:Intro}}
The high pressure phase diagram of solid hydrogen and its isotopes
is a fascinating subject of investigation during recent decades. Although
the $p-T$ phase diagrams of $ortho-H_2$ and $para-D_2$ under not too high
pressures are well understood, the same can not be said about $para-H_2$
and $ortho-D_2$ and about extremely high pressures [see, e.g., the reviews
Refs.\onlinecite{Rev.Mod.Phys.,Mao,IS}].

At low temperature and ambient pressure pure solid $o-H_2$ and $p-D_2$ consisting of molecules with orbital angular moment $J=1$ crystallize in hcp lattice with rotating molecules on the lattice sites. At lower temperature a transition takes place to the phase with the orientational long range order (LRO) of antiferroquadrupolar type. This transition is accompanied by a structural transition to fcc lattice. The transition temperature increases with pressure depending on intermolecular distance $R$ as $R^{-5}$ so indicating that it is due to the anisotropic electric quadrupole--quadrupole interaction (EQQ). This picture remains valid up to $c\sim0.55$, where $c$ is the concentration of moment bearing molecules. At intermediate concentrations ($c\gtrsim0.12$) NMR experiments have been interpreted in terms of the freezing of the orientational degrees of freedom and the transition to the quadrupolar glass phase.\cite{Rev.Mod.Phys.,Sullivan, Sullivan2} At the lower $c$ concentration there is no orientational ordering.

Molecular $p-H_2$ and $o-D_2$  also crystallize in hcp structure. However, at low pressure they remain in this structure up to $0$K. This phase has no orientational order ($J=0$) and it is called phase I (or LP  phase). At higher pressures ($\sim ${110}GPa in $p-H_{2}$  and  $\sim ${28}GPa in $o-D_{2}$) solids transform to  orientationally ordered broken symmetry
phases (phase II or BSP).
\cite{Rev.Mod.Phys.,Mao,IS}.
 The possibility of orientational order in systems of initially spherically
 symmetric molecule states is due to the involving  of higher order orbital
 moments $J=2,4...$ in the physics under pressure. The crystal field of the
 neighbors perturbs the molecular wave functions and one can gain in
 overall energy if the anisotropic EQQ interaction between nonzero
quadrupole moments is included. One has to keep in mind that EQQ
interaction rapidly increases with increasing pressure. The long range
orientational order appears abruptly at a fixed value of pressure through
the first order phase transition just as it takes place in ortho-para
mixtures when the concentration of moment bearing molecules
achieves certain fixed value.

Goncharov et al. \cite{goncharov1}
investigated the high-resolution Raman spectra of almost pure $o-D_2$.
The authors indicate that in the intermediate pressure range between the
phases I and II the ordering is incomplete and  orientational
frustration takes place. They further speculate that this intermediate
II' phase exhibits glassy behavior. Phase II' persists for a narrow
pressure range ($\sim 2$Gpa) and has abrupt boundaries.

It seems obvious that the $T-p$ phase diagram containing I, II' and II
phases can be considered in close analogy to the $T-c$ phase diagram of
ortho-para mixtures. For simplicity we imagine that $p$ maps
$c_2(p)$ -- the concentration of molecules with $J=2$, although
one should take into account other anisotropic
interactions to understand the results of the precise experiments
[e.g., Raman scattering, Ref.\onlinecite{goncharov}]. The first attempts
to describe in such a way the long range order and the orientational glass
phase in $o-D_2$ and $p-H_2$ on a microscopic theory level were done in
Refs.\onlinecite{LuTa,Schel,Schelkacheva}.

\section{$J=2$ quadrupole glass model.
\label{Sec:glass0}}

The purpose of this paper is to give a theoretical description of the possible
orientational glass transition in solid molecular para-hydrogen and
ortho-deuterium under pressure supposing that they are mixtures of $J=0$
and $J=2$ states of molecules. The theory uses the basic concepts and
methods of standard spin-glass theory. We expect our orientational glass
to correspond to the II' phase of the high pressure hydrogen phase diagram.

It is well known that the number of $J=0\rightarrow{J=2}$ transitions
increases rapidly with increasing pressure [see, e.g.,
Refs.\onlinecite{SiWi,Rev.Mod.Phys.,Mao,IS,goncharov1,goncharov,Las,Mazin,GoLo}].
The anisotropic interaction potential and the
crystal field grow rapidly with increasing density. The energy of the many--body system can be
lowered by taking advantage of the anisotropic interactions. So, the single
molecule wave functions are no longer spherical symmetric but are rather
admixtures including higher order excitations.
This admixture is probably responsible for the decrease of the
critical concentration of $J=1$ molecules for the LRO transition in
ortho-para samples at high pressure the deficit of momentum bearing
molecules being compensated by  $J=2$ molecules. It seems that the weak dependence of the
transition to the III (or A) phase on the ortho--para composition can be
described in an analogous way.

The rough estimation of $J=0\rightarrow{J=2}$ transition probability $\xi$ can be done using quantum mechanical perturbation theory considering the field of the nearest neighbours as the
perturbation:\cite{LuTa}
\begin{gather}\label{four11}
\xi= \frac{\eta} {(1+\eta)},\qquad
\eta=3.8\left(\frac{25}{26}\frac{\Gamma}{12B}\right){\left(\frac{R_{0}}{R}\right)}^{10},
\end{gather}
where $B$ is the rotational constant, $R$ is the intermolecular distance, ${\Gamma}/{B}=0.011$ in $H_{2}$ and $0.028$ in $D_{2}$.\cite{Rev.Mod.Phys.} The factor $3.8$ has a geometrical nature and it corresponds to hcp lattice. Using the
compressibility data we obtain the pressure dependence of $\xi$. This dependence is very strong. If we attribute to the probability $\xi$ the meaning of the concentration of moment bearing molecules we see that the position and the width of II' phase qualitatively coincide with that of quadrupolar glass in ortho-para mixtures. For example, $\xi=0.1$ at 40 Gpa for $o-D_{2}$. The isotope dependence is also roughly correct.

The admixture of $J=2$ states causes the frustration in the case of hcp lattice. As to disorder, it is not obvious that one can think of $J=2$ impurities as of the quenched disorder even at low temperature. Nevertheless, it seems possible to consider the whole ensemble of physically achievable realizations of mixed states as a convenient background [see, e.g., Ref.\onlinecite{Schelkachevac60}] for the formulation of an orientational glass model in the spirit of the spin-glass theory.\cite{EA}

Here we present two theoretical models of possible quadrupolar glass with $J=2$. The first
one is a generalization of well known Sherrington--Kirkpatrick\cite{sk} spin-glass and it is analogous to the model\cite{Luchinskaya} which describes well the quadrupolar glass in ortho-para
mixtures.\cite{Sullivan, Sullivan2} The second model is a generalization of the
so-called ``p-spin glass'' and it is probably more adequate for high pressures
when many particle interactions can play an important role.
We consider the case $p=3$ in detail. The results obtained for the models
differ: in the three-site model the discontinuities in the specific heat and
in the glass order parameter as the functions of the temperature do appear. We
hope that future experiments will discriminate between these models.
The essential feature of the obtained intermediate phase in both models is
the coexistence of the orientational glass with the long range orientational order
as it is seen in the experiment.\cite{goncharov1}

\section{Quadrupole glass with two--particle interaction
\label{Sec:glass}}

As the first model of the quadrupole glass we will consider
a system of particles on lattice sites $i,j$
with random truncated EQQ Hamiltonian
\begin{equation}
\hat H=-\frac{1}{2}\sum_{i\neq j}J_{ij} \hat{Q_i}\hat{Q_j}. \label{one1}
\end{equation}
Here  $J_{ij}$ are random interactions
distributed with the Gaussian probability
\begin{equation} P(J_{ij})=\frac{\sqrt{N}}{\sqrt{2\pi}
\tilde{J}}\exp\left[-\frac{(J_{ij})^{2}N}{ 2\tilde{J}^{2}}\right],
\label{two1}
\end{equation}
where the factor $N$ insures the sensible thermodynamic limit.

Operator $\hat{Q}\thicksim\left[3{J_{z}}^{2}-J(J+1)\right]$  is the axial
quadrupole moment of the hydrogen molecule in the space
$J=\mathrm{const}$; $\mathrm{Tr}\,\hat{Q}=0$. In Ref.\onlinecite{Luchinskaya}
quadrupole glass freezing in ortho--para mixtures
has been considered on the base of the Hamiltonian \eqref{one1} in the
subspace $J=1$ with $J_{z}=0,\pm1$ and $\hat{Q}=3{J_{z}}^{2}-2$, so that
\begin{equation}
\hat{Q}_{(1)}^2=2-\hat{Q}_{(1)}.\label{two11}
\end{equation}

Now the Hamiltonian \eqref{one1} will be considered in the subspace
$J=2$ with $J_{z}=0,\pm1,\pm2$ and
\begin{equation}
\hat{Q}=\frac{1}{3}\left[3{J_{z}}^{2}-6\right].\label{two112}
\end{equation}
Let us emphasize that the model with $J=2$ differs essentially from that
with $J=1$ because of the different operator algebras. For example, now
the operators $\hat Q$ are 5x5 diagonal matrices and instead of \eqref{two11} we have
\begin{equation}
\hat{Q}^3=4+4\hat{Q}-\hat{Q}^2.\label{two11111}
\end{equation}

Following the standard methods of the spin-glass theory [see, e.g.,
Ref.\onlinecite{Mezard}] and
using the replica technique, we can express the disorder-averaged free
energy of the system in the form  [see also
Refs.\onlinecite{4avtora,Schelkacheva}]
\begin{multline}\label{free}
\langle F\rangle_J/NT=\lim_{n \rightarrow 0}\frac{1}{n}\max\left \{\frac{t^2}{4}\sum_{\alpha}
(p^{\alpha})^{2} + \right.
\\
\left.\frac{t^2}{2}\sum_{\alpha>\beta} (q^{\alpha\beta})^{2}-
\ln\Tr_{\{Q^{\alpha}\}}\exp \hat{\theta}\right\},
\end{multline}
where
\begin{equation}
\hat{\theta}=t^2
\sum_{\alpha>\beta}q^{\alpha\beta}\hat{Q}^{\alpha}\hat{Q}^{\beta}+\frac{t^2}{2}
 \sum_{\alpha}{p^{\alpha}}(\hat{Q}^{\alpha})^2. \label{six}
 \end{equation}
%--------------------------------
\begin{figure}[t]
\begin{center}
\includegraphics[width=85mm]{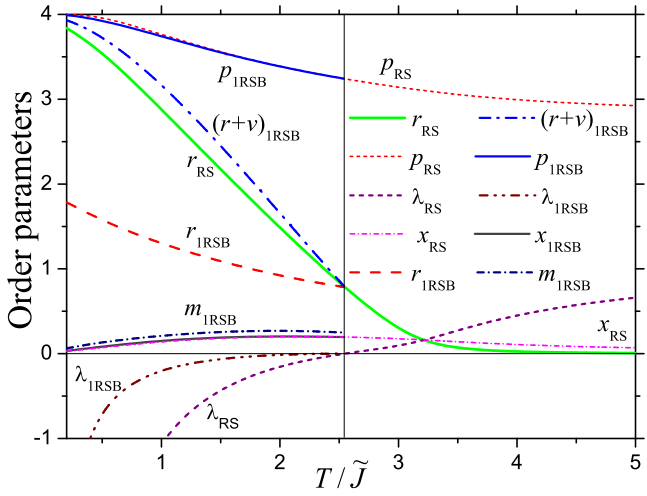} \caption{(Color online). The temperature dependence of the order
parameters for the quadrupole glass with the two--particle interaction. The
replica symmetry breaking occurs at the temperature corresponding to $\lambda_{\rm (RS)repl}=0$.}\label{fig1}
\end{center}
\end{figure}
%--------------------------------
Here $t={\tilde{J}}/T$ and the numbers $\alpha$ and $\beta$  label replicas.

The standard conditions for the free energy saddle point determine the glass order
parameter $q$ and the regular order parameter $x$ [the average quadrupole
moment],
\begin{gather}\label{four}
q^{\alpha\beta}=\langle\hat{Q}^{\alpha}\hat{Q}^{\beta}
\rangle_\theta ,
\\\label{four1}
x^{\alpha}= \langle\hat{Q}^{\alpha}\rangle_\theta ,
\end{gather}
and also the auxiliary order parameter
\begin{equation}\label{five}
p^{\alpha}= \langle(\hat{Q}^{\alpha})^2 \rangle_\theta,
\end{equation}
where
\begin{gather}
\langle\ldots\rangle_\theta=\frac{\Tr\left[(\ldots)\exp\left(\hat\theta\right)\right]}{\Tr\left[\exp\left(\hat\theta\right)\right]}.
\end{gather}

In the replica symmetric (RS) approximation, when  all $q_{\alpha
\beta }$ are equal, the expression \eqref{free} for the free energy becomes
\begin{multline}
F_{RS}=-NT\left\{ t^2\frac{q^2}{4}-t^2\frac{p^2}{4}+\overline{\ln
\Tr\left(\exp\hat{\theta}_{RS}\right)}\right\}. \label{frs}
\end{multline}
Here
\begin{gather}
\hat{\theta}_{RS}=zt\sqrt{q}\hat{Q}+t^2\frac{p-q}{2}\hat{Q}^2,
\\
\overline{(\ldots)}=\int\frac{dz}{\sqrt{2\pi}}(\ldots)\exp\left(-\frac{z^2}{2}\right)\equiv \int dz^G(\ldots).
\end{gather}

Using the extremum conditions for the free energy, Eq.\eqref{frs}, we obtain the
equations for the order parameters:
\begin{eqnarray}\label{qrs}
q&=&\overline{\langle \hat Q\rangle_{\theta_{\rm RS}}^2},
\\ \label{xrs}
x&=&\overline{\langle \hat Q\rangle_{\theta_{\rm RS}}},
\\ \label{prs}
p&=&\overline{\langle \hat Q^2\rangle_{\theta_{\rm RS}}}.
\end{eqnarray}

%--------------------------------
\begin{figure}[t]
\begin{center}
\includegraphics[height=62mm]{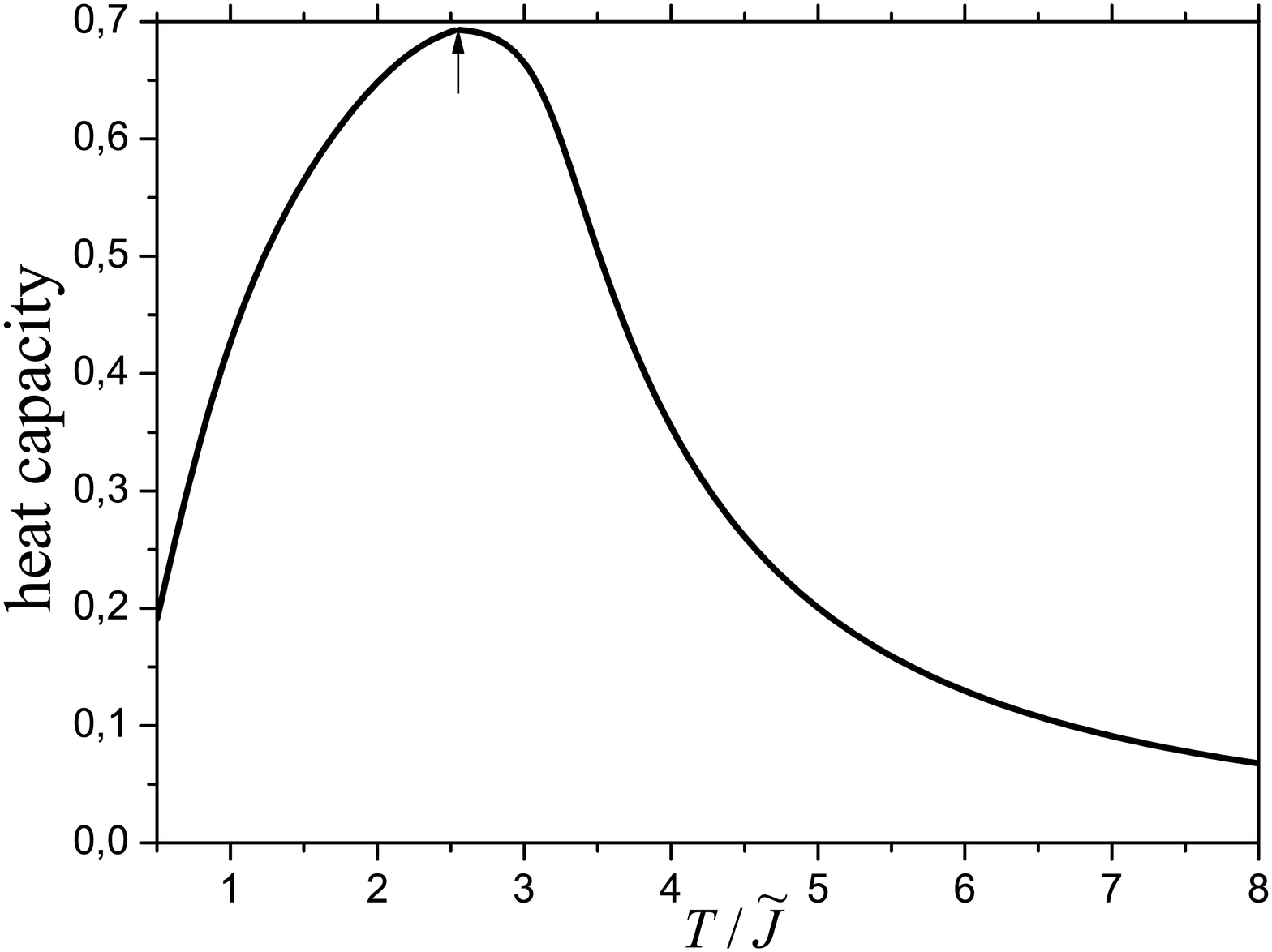} \caption{ The heat capacity of the quadrupole glass with two--particle interactions as a function of temperature. The arrow marks the temperature corresponding to the boundary between RS and more stable 1RSB solution.}\label{fig2}
\end{center}
\end{figure}
%--------------------------------
The results of numerical solution of the Eqs.(\ref{qrs}-\ref{prs}) are illustrated in Fig.\ref{fig1}.
One can see that there is the coexistence of glass and LRO. The RS glass
order parameter and the RS average quadrupole moment grow continuously on
cooling and are nonzero even at arbitrary high temperature. The absence of the
zero solution of Eqs.(\ref{qrs}-\ref{prs}) follows from the fact
that $\Tr\hat{Q}^{2k+1}\neq0$, $k=1,2,\ldots$ [see Ref.\onlinecite{4avtora} for details]. The orientational contribution to the heat capacity
\begin{gather}
\frac{C_{v(RS)}}{N}=\frac{d}{d(1/t)} \left[ t \frac{q^2-p^2}{2}\right],
\end{gather}
has a broad maximum at the temperature slightly lower than that of the instability of the RS solution. All these features are common for the quadrupole glass with $J=2$ and $J=1$ [see Refs.\onlinecite{Luchinskaya,Sullivan,Sullivan2}].

The replica symmetric solution is stable unless the replicon mode energy $\lambda$ is nonzero.\cite{Almeida} For our model we have:
\begin{gather}\label{lambda1}
\lambda_{\rm (RS) repl}= 1 - t^{2}
\overline{\left[\langle \hat Q^2\rangle_{\theta_{\rm RS}}-
(\langle \hat Q\rangle_{\theta_{\rm RS}})^2\right]^2}.
\end{gather}

At the temperature $T_0$ defined by the condition $\lambda_{\rm (RS)
repl}=0$ the RS solution becomes unstable and one needs to look for the
solutions with the broken replica symmetry (RSB).  Using the standard
procedure [see, e.g., Ref.\onlinecite{Mezard}], we perform the first stage
of the replica symmetry breaking (1RSB) according to Parisi [$n$ replicas
are divided into  $n/m$ groups with  $m$ replicas in each] and obtain the
free energy in the form [with $q^{\alpha \beta}= r_1$ if $\alpha $ and
$\beta $ are from the different groups and $q^{\alpha \beta }= r_{1}+v$ if
$\alpha $ and $\beta $ belong to the same group]
\begin{gather}\label{1frs1}
\begin{split}
F_{\mathrm{1RSB}}=&-NT\times
\\
\biggl\{& t^2\left(m \frac{r_{1}^{2}}{4}+ (1-m)\frac{(r_{1}+v)^{2}}{4}-\frac{p^2}{4}\right)+
\\
& \frac{1}{m}\int dz^G\ln\int ds^G \left[\mathrm{Tr}e^{\hat{\theta}_{1RSB}}\right]^{m}\biggr\}.
\end{split}
\end{gather}
where
\begin{gather}\label{1frs122}
\hat{\theta}_{1RSB}=zt\sqrt{{r_{1}}}\,\hat{Q}+
st\sqrt{v}\,\hat{Q}+ t^2\frac{p-r_{1}-v}{2}\hat{Q}^2.
\end{gather}
The extremum conditions for $F_{1RSB}$ yield the equations for the glass order parameters $r_{1}$ and $v$, the regular order parameter $x$, the additional order parameter $p$, and the parameter $m$ [see Appendix~\ref{ap:A}, where  $\hat{\theta}_{1RSB}$ is given by Eq.\eqref{1frs122} for $l=2$].

To estimate the form of the 1RSB solution near the bifurcation  point,  $T_0$, at which it ceases to coincide with the RS solution [i.e., in the neighborhood of $T_0$], we expand the expression for the free energy \eqref{free} up to the third order inclusively, assuming that the deviations $\delta q^{\alpha \beta}$ from $q_{RS}$ are small. In fact we expand the argument of the exponent:
\begin{multline}
\hat{\theta}=t^{2}\sum_{\alpha>\beta}\delta q^{\alpha
\beta}\hat{Q}^{\alpha}\hat{Q}^{\beta} +
\\
\frac{t^{2}}{2}p\sum_{\alpha}(\hat{Q}^{\alpha})^{2}+t^{2}q_{RS}\sum_{\alpha>\beta}\hat{Q}^{\alpha}\hat{Q}^{\beta},
\label{2frs}
 \end{multline}
with $t=t_{0}+\Delta t$.
Using the formulas of Appendix~\ref{ap:B}, we
obtain
\begin{gather}\label{10frs}
\begin{split}
&\frac{\Delta F}{NT}=\frac{t^2}{4}(1-t^{2}W)\left\{-\left[r-(m-1)v\right]^{2}+\right.
\\
&\left.v^{2}m(m-1)\right\}+\frac{t^{4}}{2}L\left[r-(m-1)v\right]^{2}-
\\
&\left.t^{6}\left\{C\left[r-(m-1)v\right]^{3}+\right.\right.
\\
&\left.D\left[r-(m-1)v\right]v^{2}m(m-1)-
\right.
\\
&\left. B_{3}v^{3}m^{2}(m-1)+B_{4}v^{3} m(m-1)(2m-1)\right\}+...
\end{split}
\end{gather}
where
$t=t_{0}+\Delta t$, $r=r_{1} - q_{RS}$, and the expressions for the parameters $W, L, C, D, B_3, B_4$ are given in Appendix \ref{ap:C}.

Using the extremum conditions for the free energy (\ref{10frs}) and the fact that $L|_{t=t_{0}}\neq {0}$,  we obtain the branching  condition $r-(m-1)v=0+o(\Delta t)^{2}$, i.e., the condition that there is  no linear term in the glass order parameters. There is no other linear term  because $(1-t^{2}W)|_{t=t_{0}}=\lambda_{\rm (RS) repl}|_{t=t_{0}}=0$
at the bifurcation point. Finally, we obtain:
 \begin{gather}\notag\label{30prs}
2\left[-\frac{t_{0}}{2}-\frac{t_{0}^{4}}{4}\frac{dW}{dt}|_{t=t_{0}}\right]\Delta
t = t_{0}^{6}\left[-B_{4}+m(-B_{3}+2B_{4})\right]v,
\\
\begin{split}
(2m-1)\biggl[-\frac{t_{0}}{2}-\frac{t_{0}^{4}}{4}&\frac{dW}{dt}|_{t=t_{0}}\biggr]\Delta
t =
\\
 t_{0}^{6}\biggl\{(2m-1)&\biggl[-B_{4}+m(-B_{3}+2B_{4})\biggr]+
\\
&m(m-1)(-B_{3}+2B_{4})\biggr\}v,
\label{40prs}
\end{split}
\end{gather}
where $ B_{3}$ and   $B_{4}$ are taken at  $T=T_0$. So,
\begin{gather}
v\sim \Delta t;\qquad\qquad  r=(m-1)v,
\label{40prs1}
\end{gather}
in the neighborhood of  $T_0$,  where the $1RSB$ solution appears and
\begin{gather}
m={\frac{B_{4}}{B_{3}}}.\label{50prs}
 \end{gather}
at the branch point $T_0$.

Let us notice that all the obtained expressions hold for Hamiltonian \eqref{one1}, where $\hat{Q}$ is the arbitrary diagonal operator such that $\Tr\hat{Q}=0 $, $\Tr\hat{Q}^{3}$ and $L|_{t=t_{0}}$ are nonzero.

For our model with  $\hat{Q}$ defined by Eq.\eqref{two112}, it follows from above formulas that $m=0.25$ and there is no jump in the order  parameters at the point where the 1RSB solution appears [as usually when $m<1$]. The solutions of the equations giving the extremum conditions of \eqref{1frs1} are presented  in Fig.\ref{fig1}. The orientational order and the glass regime coexist and grow  smoothly on cooling even through RS -- 1RSB transition. In addition, the curve for the heat capacity changes a little in passing from the RS to 1RSB solution (Fig.\ref{fig2}).

The 1RSB solution  is stable above the temperature  determined by the solution  of the condition, $\lambda_{\rm (1RSB) repl}=0$ [see Fig.\ref{fig1}],
\begin{multline}\label{lambda}
\lambda_{\rm (1RS') repl}= 1 - t^{2} \times
\\
\int dz^G \frac{\int ds^G\left[\Tr e^{\hat{\theta}_{1RSB}}\right]^{m}\left[\langle\hat Q^2\rangle_{\theta_{\rm 1RSB}}-
(\langle\hat Q\rangle_{\theta_{\rm 1RSB}})^2\right]^2}{\int ds^G
\left[\Tr e^{\hat{\theta}_{1RSB}}\right]^{m}}
\end{multline}
At $T<T_0$ we have the nonergodic state. At $T\to 0$ one expects the full replica symmetry breaking (FRSB).

\section{$p$-spin glass like quadrupole model. \label{Sec:glass1}}

In this Section we consider a generalized $p$-spin interaction spin glass
model --- $p$-quadrupole model. The 3-quadrupole case will be
considered in detail (see below). The model of Sec.\ref{Sec:glass} is 2-quadrupole model.
The $p$-spin glass model with the random interaction of $p$ Ising spins was
considered in a large number of papers [see, e.g.,
Refs.\onlinecite{Gris,Gardner,Oliveira,A.Montanari,G. Parisi1}] and serves as a generic model for
investigation of glasses without reflection symmetry.

%--------------------------------
\begin{figure}[t]
\begin{center}
\includegraphics[width=85mm]{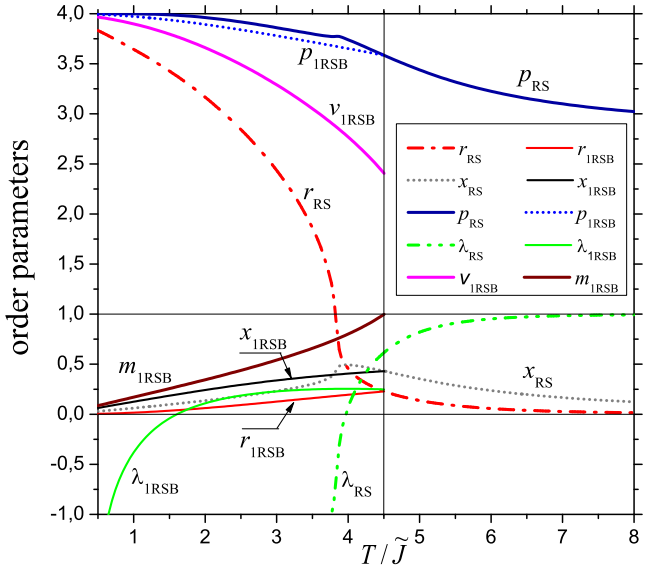} \caption{(Color online)  Order
parameters evolution with the temperature for  3-quadrupole model.  The
transition RS -- 1RSB takes place at the point defined by the
condition $m=1$. Glass order parameter $v$ has a jump at this point.}\label{fig3}
\end{center}
\end{figure}
%--------------------------------
Now we consider $l$-
quadrupole model
described by the Hamiltonian
\begin{equation}
\hat H=-\sum_{{i_{1}}\leq{i_{2}}...\leq{i_{l}}}J_{i_{1}...i_{l}}
\hat{Q}_{i_{1}}\hat{Q}_{i_{2}}...\hat{Q}_{i_{l}}, \label{one}
\end{equation}
where $i=1,2,...N$,and $\hat{Q}$ is defined in Sec.\ref{Sec:glass}.
The coupling strengths are independent random variables with a Gaussian distribution
\begin{equation}
P(J_{i_{1}...i_{l}})=\frac{\sqrt{N^{(l-1)}}}{\sqrt{l!\pi}
\tilde{J}}\exp\left[-\frac{(J_{i_{1}...i_{l}})^{2}N^{(l-1)}}{ l!\tilde{J}^{2}}\right]. \label{two}
\end{equation}

Using the replica approach we can write the free energy averaged over the disorder
[see for details Ref.\onlinecite{Gardner}] in the form:
\begin{multline}\label{free0}
\langle F\rangle_J/NT=\lim_{n \rightarrow 0}\frac{1}{n}\max\left \{- \frac{t^2}{4}\sum_{\alpha}
(p^{\alpha})^{l} + \sum_{\alpha}\mu^{\alpha}
(p^{\alpha}) +\right.
\\
\left.-\frac{t^2}{4}\sum_{\alpha\neq\beta} (q^{\alpha\beta})^{l}+ \sum_{\alpha\neq\beta}\lambda^{\alpha\beta} q^{\alpha\beta}-
\ln\Tr_{\{Q^{\alpha}\}}\exp \hat{\theta}\right\}.
\end{multline}
where
\begin{equation}\label{six1}
\hat{\theta}=
\sum_{\alpha>\beta}\lambda^{\alpha\beta}\hat{Q}^{\alpha}\hat{Q}^{\beta}+
 \sum_{\alpha}\mu^{\alpha}(\hat{Q}^{\alpha})^2.
 \end{equation}

The extremum in Eq. \eqref{free0} is taken over the physical order parameters and over the
corresponding Lagrange multipliers, $\lambda^{\alpha\beta}$ and $\mu^{\alpha}$.
So the saddle point conditions give the glass order parameter
\begin{gather}\label{four121}
q^{\alpha\beta}= \langle\hat{Q}^{\alpha}\hat{Q}^{\beta}\rangle_\theta\, ,
\end{gather}
the regular order parameter [average quadrupole moment]
\begin{gather}\label{fourR}
x^{\alpha}= \langle\hat{Q}^{\alpha}\rangle_\theta ,
\end{gather}
the auxiliary order parameter
\begin{equation}\label{five121}
p^{\alpha}= \langle(\hat{Q}^{\alpha})^2
\rangle_\theta,
\end{equation}
and the parameters
\begin{equation}\label{five11}
\lambda^{\alpha\beta}=\frac{t^2}{4}l(q^{\alpha\beta})^{(l-1)},\qquad
\mu^{\alpha}=\frac{t^2}{4}l(p^{\alpha})^{(l-1)}.
\end{equation}

Using the standard procedure [see, e.g.,
Ref.\onlinecite{Mezard}] we can obtain  RS and 1RSB  expressions for
the free energy. Let us write $F_{1RSB}$ for the case $l=3$.
The free energy $F_{RS}$ can be obtained if one put
$v=0$.

%--------------------------------
\begin{figure}[t]
\begin{center}
\includegraphics[height=60mm]{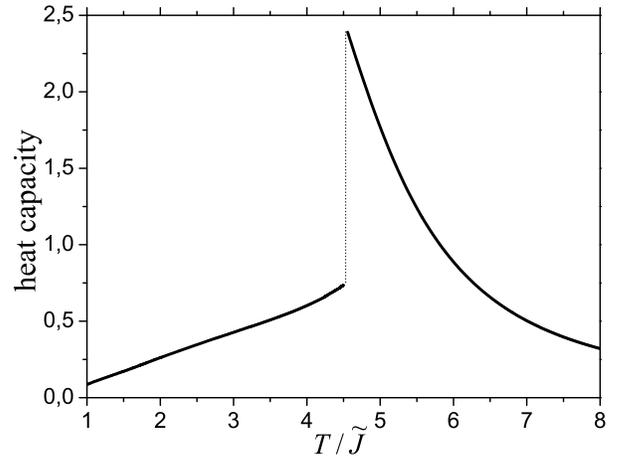} \caption{ The heat capacity of the quadrupole glass with 3--particle interactions as a function of temperature. There is a jump at the  RS -- 1RSB transition.} \label{fig4}
\end{center}
\end{figure}
%--------------------------------
\begin{gather}\label{1frs}
\begin{split}
F_{1RSB}&=-NT\times
\\
\biggl\{& m t^2\frac{r_{1}^3}{2}+ (1-m)
t^2\frac{(r_{1}+v)^3}{2}-t^2\frac{p^3}{2}+
\\
&\frac{1}{m}\int dz^G\ln
\int ds^G
\left[\Tr\exp\left(\hat{\theta}_{1RSB}\right)\right]^{m}\biggr\}.
\end{split}
\end{gather}
Here
\begin{multline}\label{1frs22}
\hat{\theta}_{1RSB}=\left.zt\sqrt{\frac{3{r_{1}}^{2}}{2}}\,\hat{Q}+\right.
\\
\left.st\sqrt{\frac{3[{(r_{1}+v)}^{2}-{r_{1}}^{2}]}{2}}\,\hat{Q}+ \right.
\\
\left. t^2\frac{3[p^{2}-{(r_{1}+v)}^{2}]}{4}\hat{Q}^2.\right.
\end{multline}

The extremum conditions for $F_{1RSB}$ yield the equations for the glass order parameters,
$r_{1}$ and $v$, the regular order parameter $x$,
the additional order parameter $p$, and the parameter $m$ [see Apendix~\ref{ap:A}, where  $\hat{\theta}_{1RSB}$ is given by Eq.\eqref{1frs22} for $l=3$].

It is easy to show that the corresponding condition $\lambda_{\rm (RS) repl}=0$ does not
determine a physical solution in the vicinity of the bifurcation point  $T_0$: one passes to the unphysical free energy branch. In fact, the transition RS -- 1RSB takes place at the point $T_{1}>T_0$ defined by the marginality conditions. At this point $m=1$ and $F_{RS}=F_{1RSB}$. There are no discontinuities in free energy. The order parameters $r$, $x$ and $p$ are continuous and $v$ has a jump [see Fig.\ref{fig3}]. When the temperature is decreased $m$ becomes smaller than one and the 1RSB solution leads to a larger (preferable) free energy than the RS solution.

The corresponding discontinuity  occurs also in the heat capacity,
\begin{gather}
\frac{C_{\rm v(1RSB)}}{N}=\frac d{d\left(\frac1t\right)}\left[ t \frac{mr_{1}^3+(1-m)(r_{1}+v)^3-p^3}{2}\right].
\end{gather}
The form of the curve for the heat capacity, Fig.\ref{fig4}, is analogous to obtained in Ref.\onlinecite{Gris} for the spherical $p$-spin model.

The 1RSB solution can be stable above the temperature $T=T_{2}$ determined by the second solution of the condition $\lambda_{\rm (1RSB)repl}=0$, see Fig.\ref{fig3},
\begin{multline}\label{lambda121}
\lambda_{ (\mathrm{1RSB})\rm repl}= 1 - t^{2} \frac{l(l-1)(r_{1}+v)^{(l-2)}}{2} \times
\\
\int dz^G \frac{\int ds^G\left[\Tr e^{\hat{\theta}_{1RSB}}\right]^{m}\{\langle\hat Q^2\rangle_{\theta_{\rm 1RSB}}-
(\langle\hat Q\rangle_{\theta_{\rm 1RSB}})^2\}}{\int ds^G
\left[\Tr e^{\hat{\theta}_{1RSB}}\right]^{m}}
\end{multline}
At the point $T_2$ a transition to FRSB-state or to a stable 2RSB-state may take place.

\section{Conclusions \label{Sec:Conc}}

In this paper we give a theoretical description of possible
orientational glass transition in solid molecular para-hydrogen and
ortho-deuterium under pressure supposing that they are mixtures of $J=0$
and $J=2$ states of molecules. The theory uses the basic concepts and
methods of the standard spin-glass theory. We expect that our orientational glass
corresponds to the II' phase of the high pressure hydrogen phase diagram.

We present two theoretical models of possible quadrupole glass with $J=2$. The first
one is a generalization of thewell known Sherrington--Kirkpatrick spin-glass.
The second model is a generalization of so-called ``$p$-spin glass'' and it is probably more adequate for high pressures when many particle interactions can play an important role. We consider in detail the case $p=3$. The results obtained for two models differ: in three-site model the discontinuities in the heat capacity and in the glass order parameter as functions of the temperature do appear. We hope that future experiments will discriminate between these models. The essential feature of the obtained intermediate phase in both models is the coexistence of the orientational glass with the long range orientational order as it is seen experimentally.\cite{goncharov1}

\section{Acknowledgments}
Authors thank V.N. Ryzhov for
helpful discussions and valuable comments.

This work was supported in part by the Russian Foundation for Basic Research (Grant
No.08-02-00781; by the President of the Russian Federation (Grant No. 07-02-00998MK),
the Russian  Foundation for National Science Support, the Dynasty Foundation, the Russian Academy of Sciences program ``Quantum Macrophysics'' and the Presidium of Russian Academy of Sciences program.

\appendix

\section{\label{ap:A}}

The equations  for 1RSB  glass order parameters $r_{1}$ and $v$, the regular order parameter $x$, the
additional order parameter $p$, and the parameter $m$.
\begin{gather}\label{1qrs}
\begin{split}
r_{1}&=
\\
&\int_{z^G}\left\{ \frac{\int_{s^G}{\left[\Tr e^{\hat{\theta}_{\rm 1RSB}}\right]}^{(m-1)}\left[\Tr\hat{Q} e^{\hat{\theta}_{\rm 1RSB}}\right]}
{\int_{s^G}{\left[\Tr e^{\hat{\theta}_{\rm 1RSB}}\right]}^{m}}\right\}^{2},
\end{split}
\\\label{1vrs}
v+r_{1}=
\int_{z^G}\frac{\int_{s^G}{\left[\Tr e^{\hat{\theta}_{\rm 1RSB}}\right]}^{(m-2)}{\left[\Tr{\hat{Q} }e^{\hat{\theta}_{\rm 1RSB}}\right]}^{2}}
{\int ds^G{\left[\Tr e^{\hat{\theta}_{\rm 1RSB}}\right]}^{m}},
\\\label{1xrs}
x=\int_{z^G} \frac{\int_{s^G}{\left[\Tr e^{\hat{\theta}_{\rm 1RSB}}\right]}^{(m-1)}\left[\Tr{\hat{Q} } e^{\hat{\theta}_{\rm 1RSB}}\right]}
{\int_{s^G}{\left[\Tr e^{\hat{\theta}_{\rm 1RSB}}\right]}^{m}},
\\\label{1prs}
p=\int_{z^G} \frac{\int_{s^G}{\left[\Tr e^{\hat{\theta}_{\rm 1RSB}}\right]}^{(m-1)}\left[\Tr{\hat{Q} }^{2} e^{\hat{\theta}_{\rm 1RSB}}\right]}
{\int_{s^G}{\left[\Tr e^{\hat{\theta}_{\rm 1RSB}}\right]}^{m}},
\end{gather}
and
\begin{widetext}
\begin{gather}\label{1mrs}
%\begin{split}
m\frac{t^{2}}{4}(l-1)
%&
\bigl[{(r_{1}+v)}^{l}-{(r_{1})}^{l}\bigr]=
%\\&
-\frac{1}{m}\int_{z^G}\ln \int_{s^G}\left[\Tr e^{\hat{\theta}_{1RSB}}\right]^{m}+
%\\&
\int_{z^G} \frac{\int_{s^G}{\left[\Tr e^{\hat{\theta}_{1RSB}}\right]}^{m}\ln\left[\Tr e^{\hat{\theta}_{1RSB}}\right]}
{\int_{s^G}{\left[\Tr e^{\hat{\theta}_{1RSB}}\right]}^{m}}.
%\end{split}
\end{gather}
%\end{widetext}

\section{\label{ap:B}}

The only nonzero sums are
\begin{gather}\label{11eq:JP}
\lim_{n \rightarrow 0}\frac{1}{n}
{\sum}'_{\alpha,\beta}(\delta
q^{\alpha\beta})^{3}=(m-1)\eta^{3}-m\xi^{3};
\end{gather}
and
%\begin{widetext}
\begin{gather}
%\begin{split}
\lim_{n \rightarrow
0}\frac{1}{n}{\sum}'_{\alpha,\beta,\gamma}\delta
q^{\alpha\beta}
%&
\delta q^{\beta\gamma}\delta
q^{\gamma\alpha}=(m-1)(m-2)\eta^{3} -
%\\&
3m(m-1)\eta\xi^{2}+2m^{2}\xi^{3};
%\end{split}
\\
%\begin{split}
\lim_{n \rightarrow
0}\frac{1}{n}{\sum}'_{\alpha,\beta,\gamma}
%&
(\delta
q^{\alpha\beta})^{2}\delta
q^{\alpha\gamma}=(m-1)^{2}\eta^{3}-
%\\& 
m(m-1)\left(\eta\xi^{2}+\eta^{2}\xi\right)+m^{2}\xi^{3};
%\end{split}
\\
%\begin{split}
\lim_{n \rightarrow 0}
%&
\frac{1}{n}{\sum}'_{\alpha,\beta,\gamma,\delta} \delta
q^{\alpha\beta}\delta q^{\alpha\gamma}\delta q^{\beta\delta}=
%\\&
\quad(m-1)^{3}\eta^{3}+
3m^{2}(m-1)\eta\xi^{2}-
%\\&
\quad 3m(m-1)^{2}\eta^{2}\xi-m^{3}\xi^{3},
%\end{split}
\end{gather}
%\end{widetext}
where $\eta=r+v$ and $\xi=r$.  The prime on the sum means that only the superscripts belonging to the same
$\delta q$ are necessarily different in  ${\sum}'$.

\section{\label{ap:C}}

The formulas used to calculate the parameters are
\begin{eqnarray}
W&=&\langle\hat{Q}_{1}^2\hat{Q}_{2}^2\rangle-2\langle\hat{Q}_{1}^2\hat{Q}_{2}\hat{Q}_{3}\rangle+\langle\hat{Q}_{1}\hat{Q}_{2}\hat{Q}_{3}\hat{Q}_{4}\rangle;
\\
L&=&-\langle\hat{Q}_{1}^2\hat{Q}_{2}\hat{Q}_{3}\rangle+\langle\hat{Q}_{1}\hat{Q}_{2}\hat{Q}_{3}\hat{Q}_{4}\rangle;
\\
C&=&-(B_{2}+B'_{2})+2B_{3}+B'_{3}-B_{4};
\\
D&=&-3B_{3}-B'_{3}+3B_{4};
\end{eqnarray}
where
%\begin{widetext}
\begin{eqnarray*}
B_{2}&=&\frac{1}{2}\langle\hat{Q}_{1}^{2}\hat{Q}_{2}^{2}\hat{Q}_{3}\hat{Q}_{4}\rangle+
\frac{1}{2}\langle\hat{Q}_{1}\hat{Q}_{2}\hat{Q}_{3}\hat{Q}_{4}\hat{Q}_{5}\hat{Q}_{6}\rangle-
 \langle\hat{Q}_{1}^{2}\hat{Q}_{2}\hat{Q}_{3}\hat{Q}_{4}\hat{Q}_{5}\rangle;
\\
B'_{2}&=&\frac{1}{3}\langle\hat{Q}_{1}\hat{Q}_{2}\hat{Q}_{3}\hat{Q}_{4}\hat{Q}_{5}\hat{Q}_{6}\rangle-
\frac{1}{2}\langle\hat{Q}_{1}^{2}\hat{Q}_{2}\hat{Q}_{3}\hat{Q}_{4}\hat{Q}_{5}\rangle+\frac{1}{6}\langle\hat{Q}_{1}^{3}\hat{Q}_{2}\hat{Q}_{3}\hat{Q}_{4}\rangle;
\\
B_{3}&=&\frac{1}{6}\langle\hat{Q}_{1}^{2}\hat{Q}_{2}^{2}\hat{Q}_{3}^{2}\rangle-
\frac{1}{2}\langle\hat{Q}_{1}^{2}\hat{Q}_{2}^{2}\hat{Q}_{3}\hat{Q}_{4}\rangle-
\frac{1}{6}\langle\hat{Q}_{1}\hat{Q}_{2}\hat{Q}_{3}\hat{Q}_{4}\hat{Q}_{5}\hat{Q}_{6}\rangle+
 \frac{1}{2}\langle\hat{Q}_{1}^{2}\hat{Q}_{2}\hat{Q}_{3}\hat{Q}_{4}\hat{Q}_{5}\rangle;
\\
B'_{3}&=&-\langle\hat{Q}_{1}\hat{Q}_{2}\hat{Q}_{3}\hat{Q}_{4}\hat{Q}_{5}\hat{Q}_{6}\rangle+
 \frac{5}{2}\langle\hat{Q}_{1}^{2}\hat{Q}_{2}\hat{Q}_{3}\hat{Q}_{4}\hat{Q}_{5}\rangle-
 \frac{1}{2}\langle\hat{Q}_{1}^{3}\hat{Q}_{2}\hat{Q}_{3}\hat{Q}_{4}\rangle-\frac{3}{2}\langle\hat{Q}_{1}^{2}\hat{Q}_{2}^{2}\hat{Q}_{3}\hat{Q}_{4}\rangle+\frac{1}{2}\langle\hat{Q}_{1}^{3}\hat{Q}_{2}^{2}\hat{Q}_{3}\rangle;
\\
B_{4}&=&\frac{1}{3}\langle\hat{Q}_{1}\hat{Q}_{2}\hat{Q}_{3}\hat{Q}_{4}\hat{Q}_{5}\hat{Q}_{6}\rangle-
 \langle\hat{Q}_{1}^{2}\hat{Q}_{2}\hat{Q}_{3}\hat{Q}_{4}\hat{Q}_{5}\rangle+
\frac{1}{3}\langle\hat{Q}_{1}^{3}\hat{Q}_{2}\hat{Q}_{3}\hat{Q}_{4}\rangle+\frac{3}{4}\langle\hat{Q}_{1}^{2}\hat{Q}_{2}^{2}\hat{Q}_{3}\hat{Q}_{4}\rangle
-\frac{1}{2}\langle\hat{Q}_{1}^{3}\hat{Q}_{2}^{2}\hat{Q}_{3}\rangle+
 \frac{1}{12}\langle\hat{Q}_{1}^{3}\hat{Q}_{2}^{3}\rangle;
\end{eqnarray*}
and
\begin{eqnarray}
\langle\hat{Q}_{\gamma}^k\hat{Q}_{\delta}^n...\rangle &=&
\frac{\Tr\left[(\hat{Q}^{\gamma})^k(\hat{Q}^{\delta})^n... \exp\Xi\right]}
{\Tr\left[\exp\Xi\right]} \label{0prs},
\\
\hat{\Xi}&=&\frac{t^{2}}{2}p\sum_{\alpha}(\hat{Q}^{\alpha})^{2}+t^{2}q_{RS}\sum_{\alpha>\beta}\hat{Q}^{\alpha}\hat{Q}^{\beta}.
\end{eqnarray}
\end{widetext}


\begin{thebibliography}{99}

\bibitem{Rev.Mod.Phys.} L.F. Silvera, Rev. Mod. Phys.  \textbf{52}, 393 (1980).

\bibitem{Mao} H.K. Mao and R.J. Hemley, Rev. Mod. Phys.  \textbf{66}, 671 (1994).

\bibitem{IS} I.F. Silvera, Proc. of Joint 20th AIRAPT and 43th EHPRG International
Conference on High Pressure Science and Technology, (Karlsruhe, Germany, June 27-July 1, 2005), edited by E. Dinjus and N. Dahmen (Forschungszentrum Karlsruhe GmbH, Karlsruhe, 2005) PL-O217.


\bibitem{Sullivan}  N.S. Sullivan, M. Devoret, B.P. Cowan and C. Urbina, Phys.
Rev. B\textbf{17}, 5016 (1978).

\bibitem{Sullivan2}  N.S. Sullivan, C.M. Edwards and J.R. Brookeman, Mol. Cryst. Liq. Cryst. \textbf{ 139}, 385(1986).

\bibitem{goncharov1} A.F. Goncharov, J.H. Eggert, I.I. Mazin, R.J. Hemley, and H.K.Mao, Phys. Rev. B \textbf{54}, R15590(1996).

\bibitem{goncharov} A.F. Goncharov, M.A. Strzhemechny, H.K. Mao, and R.J. Hemley,
Phys. Rev. B \textbf{63}, 064304 (2001).

\bibitem{LuTa} E.A. Lutchinskaia, and E.E. Tareyeva, in Proc. of the
Conference on the High Pressure Effects in Materials, (Kiev, 1986), p.21.


\bibitem{Schel} T.I. Schelkacheva, Phys. Lett. A \textbf{239}, 397(1998).

\bibitem{Schelkacheva} T.I. Schelkacheva, JETP. Lett. \textbf{76}, 374
(2002) [Pis'ma Zh. \'{E}ksp. Teor. Fiz. \textbf{76}, 434 (2002)].

\bibitem{SiWi} I.F. Silvera and R.J. Wijngaarden, Phys. Rev. Lett. \textbf{ 47},
39(1981).

\bibitem{Las} L. Lassche, I.F. Silvera and A. Lagendijk, Phys. Rev. Lett. \textbf{65}, 2677 (1990).

\bibitem{GoLo}I. Goncharenko, and P. Loubeyre, Nature, \textbf{435}, 1206
(2005).

\bibitem{Mazin} A.F. Goncharov, J.H. Eggert, I.I. Mazin, R.J. Hemley, and H.K. Mao,
Phys. Rev. B \textbf{54}, R15590 (1996).

\bibitem{Schelkachevac60} T.I. Schelkacheva, E.E. Tareyeva, and N.M. Chtchelkatchev,
 Phys. Rev. B \textbf{76}, 195408 (2007).

\bibitem{EA} S.F. Edwards, and P.W. Anderson, J. Phys. F {\bf 5}, 965 (1975).

\bibitem{sk} D. Sherrington, and S. Kirkpatrick, Phys. Rev. Lett. {\bf
32}, 1972 (1975); S. Kirkpatrick, and D. Sherrington, Phys. Rev. B {\bf 17},
4384 (1978).

\bibitem{Luchinskaya} E.A. Lutchinskaya, V.N. Ryzhov, and E.E. Taryeva, J. Phys. C \textbf{17}, L665
(1984); E.A. Lutchinskaia, and E.E. Tareyeva, Phys.Rev. B \textbf{52}, 366 (1995).

\bibitem{Mezard} M. Mezard, G. Parisi, and M. Virasoro, Spin Glass Theory and
beyond (World Scientific, Singapore, 1987).

\bibitem{4avtora} {N.V. Gribova, V.N. Ryzhov, T.I. Schelkacheva,
E.E. Tareyeva}, Phys. Lett. A \textbf{315}, 467( 2003).

\bibitem{Almeida} J.R.L.Almeida and D.J.Tauless, J.Phys.A {\bf 11}, 983(1978).

\bibitem{Gardner} E. Gardner,  Nuc. Phys. {\bf B257}, 747 (1985).
\bibitem{Gris} A. Crisanti and H.-J. Sommers, Z. Phys. B {\bf87}, 341(1992).

\bibitem{G. Parisi1} M. Campellone, B. Coluzzi, and G. Parisi,
 Phys. Rev. B \textbf{58}, 12081 (1998).

\bibitem{Oliveira} V. M. de Oliveira and J.F.Fontanari,
 J. Phys. A: Math. Gen. \textbf{32}, 2285 (1999).

\bibitem{A.Montanari} A. Montanari, and F. Ricci-Tersenghi, cond-mat/0301591,
Eur. Phys.J. B {\bf 33}, 339(2003).

\end{thebibliography}
\end{document}